\documentclass[a4paper]{article}

\usepackage{float}
\usepackage{subcaption}
\usepackage{multirow} 
\newcommand{\tabincell}[2]{\begin{tabular}{@{}#1@{}}#2\end{tabular}} 
\usepackage{graphicx}
\usepackage{tabularx}

\newcommand\clearrow{\global\let\rowmac\relax}
\usepackage{caption} 
\renewenvironment{thebibliography}[1]{
  \begin{oldthebibliography}{#1}
  \setlength{\itemsep}{0.em}
  \setlength{\parskip}{0.em}
}
{
  \end{oldthebibliography}
}

\usepackage{INTERSPEECH2021}

\title{Layer-wise Fast Adaptation for End-to-End Multi-Accent Speech Recognition}

\name{Xun Gong, Yizhou Lu, Zhikai Zhou, Yanmin Qian$^\dagger$
\thanks{
    $^\dagger$corresponding author
}
}
\address{
MoE Key Lab of Artificial Intelligence, AI Institute \\
X-LANCE Lab, Department of Computer Science and Engineering \\
Shanghai Jiao Tong University, Shanghai, China
}
\email{\{gongxun,luyizhou4,zhikai.zhou,yanminqian\}@sjtu.edu.cn}

\begin{document}
\maketitle

\begin{abstract}
Accent variability has posed a huge challenge to automatic speech recognition~(ASR) modeling. Although one-hot accent vector based adaptation systems are commonly used, they require prior knowledge about the target accent and cannot handle unseen accents. Furthermore, simply concatenating accent embeddings does not make good use of accent knowledge, which has limited improvements. In this work, we aim to tackle these problems with a novel layer-wise adaptation structure injected into the E2E ASR model encoder. The adapter layer encodes an arbitrary accent in the accent space and assists the ASR model in recognizing accented speech. Given an utterance, the adaptation structure extracts the corresponding accent information and transforms the input acoustic feature into an accent-related feature through the linear combination of all accent bases. We further explore the injection position of the adaptation layer, the number of accent bases, and different types of accent bases to achieve better accent adaptation. Experimental results show that the proposed adaptation structure brings 12\% and 10\% relative word error rate~(WER) reduction on the AESRC2020 accent dataset and the Librispeech dataset, respectively, compared to the baseline.
\end{abstract}


\section{Introduction}

\label{sec:introduction}


In recent years, end-to-end (E2E) automatic speech recognition~(ASR) models, which directly optimize the probability of the output sequence given input acoustic features, have made great progress in a wide range of speech corpora~\cite{comparative_study_transformer_rnn_2019}. 
One of the most pressing needs for ASR today is the support for multiple accents in a single system, which is often referred to as multi-accent speech recognition in the literature.
The difficulties of recognizing accented speech, including phonology, vocabulary and grammar, have posed a serious challenge to current ASR systems~\cite{AESRC2020Datasetshi}.
A straightforward method is to build a single ASR model from mixed data (accented speech from non-native speakers and standard data from native speakers). 
However, such models usually suffer from severe performance degradation due to the accent mismatch during training and inference~\cite{chen2015improving,elfeky2016towards,li2018multi,jain2018improved,Grace_Bastani_Weinstein_2018}.
Previous work has explored different accent adaptation methods for acoustic models.
MixNet~\cite{jain2018improved,jainMultiAccentAcousticModel2019} is based on Mixture of Experts~(MoE) architecture, where experts are specialized to segregate accent-specific speech variabilities. 
Model-agnostic meta-learning~(MAML)~\cite{winata2020learning} approach is also explored to learn the rapid adaptation to unseen accents.
One-hot accent vectors are well utilized to build multi-basis adaptation~\cite{Grace_Bastani_Weinstein_2018,yang2018joint,yoo2019highly}, where each basis is designed to cover certain types of accents.
Recently, with the improvements in accent identification~(AID) models~\cite{huangAISPEECHSJTUACCENTIDENTIFICATION}, several methods have been explored to integrate AID into speech recognition for accent adaptation.
For example, \cite{jain2018improved,turan2020achieving} proposed to concatenate accent embeddings and acoustic features for adaptation of the acoustic model. \cite{jain2018improved,jainMultiAccentAcousticModel2019} proposed a multi-task framework to jointly model both ASR and AID tasks.
Meanwhile, these approaches have also been applied to the E2E ASR framework~\cite{li2018multi,viglino2019end}.



In this paper, we study a novel approach to rapid adaptation of accented data via the layer-wise transformation of input features.
Compared to previous works, the proposed method stimulates the potential of both accent embeddings and hidden representations.
Instead of simply concatenating accent embeddings and input features, we adopt a different scheme with scaling and shifting transformations, which has been proven a valuable method to utilize the accent embeddings~\cite{Grace_Bastani_Weinstein_2018,tanAISPEECHSJTUASRSYSTEM,yoo2019highly}. 
Furthermore, we propose the multi-basis adapter layer architecture to represent the accent-dependent features.
The adapter basis based approach has shown its potential in various fields, including computer vision~\cite{adapter_cv_2017}, natural language processing~\cite{adapter_nlp_2019}, neural machine translation~\cite{adapter_nmt_2019} and multi-lingual ASR~\cite{kannanAdapterRNNT2019}.
Similarly, multiple bases are also proved to be effective in speaker adaptation~\cite{tanCAT2015,wuWuchunyangIvector12016} and code-switching ASR task~\cite{luMoELuyizhou2020}.
However, the effectiveness of such approaches in multi-accent speech recognition has not been investigated to the best of our knowledge.
In this paper, we incorporate the adapter basis based technique into the E2E ASR architecture for multi-accent speech recognition.
Furthermore, we downsize the typically massive bases to much smaller modules in each adapter layer.
As the proposed method models different accents in the continuous embedding space, it can naturally cope with unseen accents in the inference stage by a linear combination of adapter bases.
During adaptation, interpolation coefficients between different adapter bases are predicted from the accent embeddings.
With the proposed framework, accent adaptation can be achieved in a parameter-efficient and flexible manner.



The rest of the paper is organized as below: 
In Section~\ref{sec:methods}, we present our layer-wise adapter architecture with the multi-task regularization.
Experimental results are presented and analyzed in Section~\ref{sec:experiments}.
Finally, the conclusion is given in Section~\ref{sec:conclusions}.

\section{Layer-wise Fast Adaptation on E2E Multi-Accent ASR}
\label{sec:methods}
In this section, we first give a brief review of the joint connectionist temporal classification~(CTC)-attention based E2E ASR. 
Then we describe the proposed accent adapter layer and corresponding training strategies. 
The new approach mainly includes two parts: the adapter layer construction and interpolation coefficients regularization.

\def \x {\mathbf{x}}
\def \y {{\mathbf{y}}}
\def \h {\mathbf{h}}
\def \a {\mathbf{a}}
\def \W {\mathbf{W}}

\newcommand{\adapt}{\mathcal{A}}
\newcommand{\hin}{\boldsymbol{h^i}}
\newcommand{\hac}{\boldsymbol{h^o}}
\newcommand{\zz}{\boldsymbol{z}}
\newcommand{\al}{\boldsymbol{\alpha}}
\newcommand{\hal}{\boldsymbol{\alpha^{\text{(ref)}}}}
\newcommand{\alk}{\alpha_k}
\newcommand{\ai}[1]{\alpha_#1}

\begin{figure*}
\centering
\includegraphics[width=\linewidth]{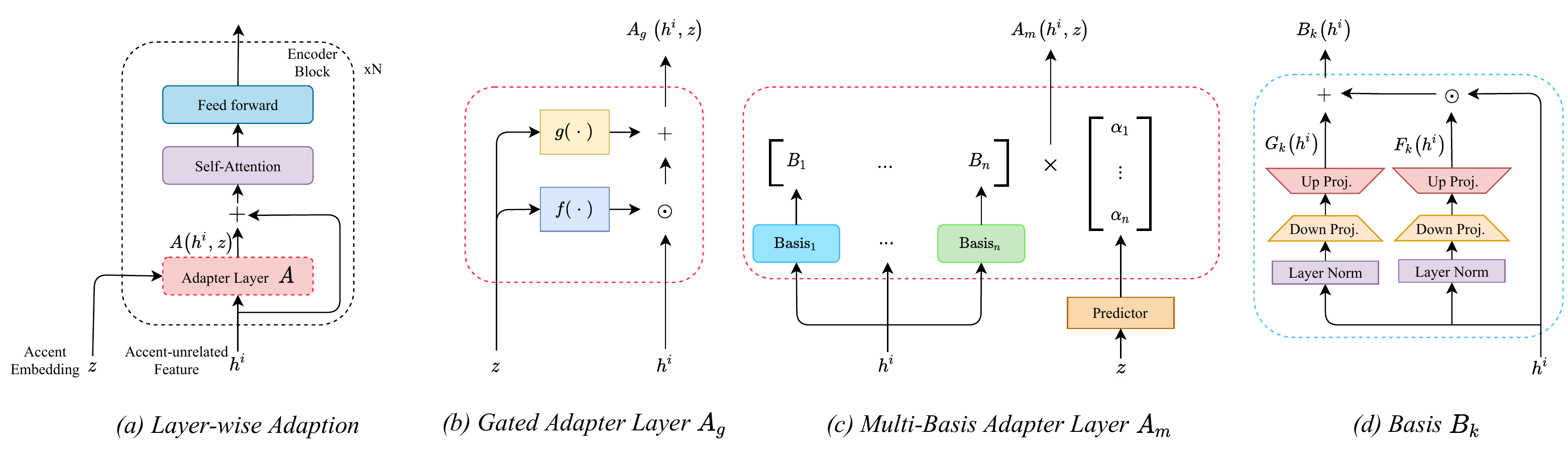}
\caption{Schematic diagram of the proposed adapter layer. The adapter layer in (a) is optionally inserted in each encoder block, which is discussed in Section~\ref{sec:expr_position}. Here, $+$, $\times$, and $\odot$ denote summation, matrix multiplication, and element-wise product, respectively.}
\label{fig:pipeline}
\end{figure*}

\subsection{Pretrained transformer-based E2E ASR}
\label{sec:baseline}

The transformer is a sequence-to-sequence~(S2S) structure consisting of a multi-layer encoder and a multi-layer decoder~\cite{attention_is_all_you_need}.
The encoder takes acoustic features as input to be mapped into high-level representations $\h$.
The decoder network utilizes the encoded representation $\h$ with an attention mechanism and outputs the predicted tokens auto-regressively.
At each decoding step, the decoder emits the posterior probabilities of the next token given previous outputs.
We train the transformer model with the joint CTC-attention framework~\cite{JointCTCAttentionBased_kim_2017} to exploit the advantages from both CTC and attention-based models. 
The loss function is defined as below:
\begin{equation}
\label{eq:joint_ctc}
\begin{aligned}
\mathcal{L}_{jca} = \lambda_{ctc} \mathcal{L}_{ctc} + (1 - \lambda_{ctc}) \mathcal{L}_{s2s}
\end{aligned}
\end{equation}
where $\mathcal{L}_{ctc}$ and $\mathcal{L}_{s2s}$ are the CTC and S2S objective losses, respectively.
A tunable parameter $\lambda_{ctc} \in [0,1]$ is used to control the contribution of each loss.



\subsection{Adapter Layer}
\label{sec:layer}

The E2E ASR model trained on common standard corpora usually lacks generalization on accented data due to the accent mismatch. 
Adapter layers are injected into the ASR encoder blocks to transform the accent-unrelated features into the accent-related space.
The architecture of the new ASR encoder with the proposed adapter layer is illustrated in Figure~\ref{fig:pipeline}(a).
The adapter layer, hereinafter denoted as $\adapt$, is used as a pre-processing to transform accent-unrelated features into accent-related features.
Denote by $\hin$ the input feature before the encoder block, $\zz$ the accent embedding, and $\adapt(\hin, \zz)$ the output feature in the accent-related space.
The output feature $\adapt(\hin, \zz)$ is then wrapped into the encoder block by a residual connection ($+$) as shown in Figure~\ref{fig:pipeline}(a), to enable the original acoustic information to flow through later encoder layers.
Different types of adapter layers $\adapt$ are explored in the following sections: $\adapt_g$ in Section~\ref{sec:ebd} and $\adapt_m$ in Section~\ref{sec:multi_basis}.

\subsubsection{Gated Adapter Layer}
\label{sec:ebd}

The first scheme to attain the transform function follows our previous investigation in~\cite{tanAISPEECHSJTUASRSYSTEM}. 
As shown in Figure~\ref{fig:pipeline}~(b), a scaling factor $f(\zz)$ and a shifting factor $g(\zz)$ can be applied to the input feature for accent adaptation:
{\setlength\abovedisplayskip{4pt}
\setlength\belowdisplayskip{4pt}
\begin{equation}
\label{eq:ebd}
\adapt_g(\hin, \zz) = f(\zz) \odot \hin + g(\zz)
\end{equation}
}where $\adapt_g$ is the gated adapter layer, and $\odot$ denotes the element-wise product. $f(\zz)$ and $g(\zz)$ are separately generated by a single dense layer with $\mathrm{tanh}(\cdot)$ activation.

\subsubsection{Multi-basis Adapter Layer}
\label{sec:multi_basis}

The second scheme is to build a multi-basis adapter layer as in Figure~\ref{fig:pipeline}~(c).
The multi-basis layer concatenates the output $B_k(\hin)$ from each basis with the corresponding interpolation coefficient $\alk$. 
Similar to Section~\ref{sec:ebd}, the scaling $F_k(\cdot)$ and shifting $G_k(\cdot)$ modules are used to transform the input $\hin$ into the accent-related space as shown in Figure~\ref{fig:pipeline}~(d), where $k = 1, 2, \ldots, n$ and $n$ is the number of adapter bases.
{\setlength\abovedisplayskip{2pt}
\setlength\belowdisplayskip{2pt}
\begin{equation}
\label{eq:adapter_layer}
\begin{aligned}
\adapt_m(\hin, \zz) = &\sum_{k=1}^n \ai{k} B_k(\hin) \\
                    = &\sum_{k=1}^n \ai{k} * \{ F_k(\hin) \odot \hin + G_k(\hin) \}
\end{aligned}
\end{equation}
}Note that one can also use scaling-only $\left(G_k(\hin)=\mathbf{0}\right)$ and shifting-only $\left(F_k(\hin)=\mathbf{0}\right)$ operations in the bases, which will be discussed in Section~\ref{sec:expr_type}.

\noindent\textbf{Projection Module}\quad To make the bases in Figure~\ref{fig:pipeline}~(d) simple and flexible, we propose a sandglass-style structure for $F(\cdot)$ and $G(\cdot)$ modeling: a down-projection network and a up-projection network with the non-linear activation $\mathrm{ReLU}(\cdot)$.
This architecture allows us to easily adjust the modules' capacity, depending on the complexity of the accents.
Additionally, we normalize the input of each adapter basis by LayerNorm~\cite{ba2016layernorm}.

\noindent\textbf{Predictor}\quad Different from the one-hot accent vector that is commonly used in prior works on accent adaptation~\cite{Grace_Bastani_Weinstein_2018,yang2018joint}, here we adopt a soft assignment of bases by interpolating between all adapter bases dynamically.
To estimate interpolation coefficients $\al \in \mathbb{R}^n$ from accent embedding $\zz$, a predictor $p(\cdot)$ model is used, and give guidance on the usage of modules.
\begin{align}
\label{eq:predictor}
\al = \operatorname{SoftMax} (p(\zz)) \text{, where } 1 = \sum_{k=1}^n \alk
\end{align}
where the interpolation coefficients $\al = (\ai{1}, \ldots , \ai{n})$ are probabilities for multiple bases.
The predictor $p(\cdot)$ can be composed of several DNN layers.


\subsubsection{Multi-task Regularization}
\label{sec:multi_task}


During training, we found that, without any constraints, the distribution of interpolation coefficients $\al$ would rapidly reduce to a certain basis for all accents, which greatly limits the adapter layer's adaptation capability.
Thus, we apply the multi-task learning~(MTL) scheme to utilize the loss from an auxiliary task, i.e.~the predictor in Section~\ref{sec:multi_basis}, to regularize the training of both ASR and predictor models.
An auxiliary loss from the predictor is introduced to the ASR loss $\mathcal{L}_{jca}$, and then the final loss $\mathcal{L}_{mtl}$ for the entire system is calculated as:
{\setlength\abovedisplayskip{3pt}
\setlength\belowdisplayskip{3pt}
\begin{equation}
\label{eq:mtl_loss}
\begin{aligned}
\mathcal{L}_{mtl} = \mathcal{L}_{jca} + \gamma_{mtl} \mathcal{L}_{\text{MSE}} (\hal, \al)
\end{aligned}
\end{equation}
}where $\hal$ is the target label of the predictor outputs $p(\zz)$, $\al$ is the output of the predictor, and $\gamma_{mtl}$ is a hyperparameter to control the contribution of the predictor loss.
The target label $\hal$ is obtained via the clustering of accent embeddings extracted from the pretrained AID model.
The number of clusters is set to $n$, and here the K-means algorithm is adopted.

\section{Experiments}
\label{sec:experiments}

\subsection{Setup}
\label{sec:expr_setup}

\subsubsection{Dataset}
\label{sec:expr_dataset}

Our experiments are conducted on the Accented English Speech Recognition Challenge 2020~(AESRC2020) dataset~\cite{AESRC2020Datasetshi} and the Librispeech corpus~\cite{panayotovLibrispeechASRCorpus2015}. 
AESRC2020 contains a training set for 8 English accents in England~(UK), America~(US), China~(CHN), Japan~(JPN), Russia~(RU), India~(IND), Portugal~(PT) and Korea~(KR), with 20-hour data for each accent, however two more accents Canada~(CAN) and Spain~(ES) are included in test set while cv set has eight accents. 
Librispeech contains a 960-hour training set, while dev-clean/other~(dev~c/o) and test-clean/other~(test~c/o) are used for standard tests.
We report the word error rate~(WER) on all evaluation sets.

\subsubsection{E2E based Baseline}
\label{sec:expr_baseline}

For acoustic feature extraction, 80-dimensional fbank features are extracted with a step size of 10ms and a window size of 25ms, and utterance-level cepstral mean and variance normalization (CMVN) is applied on the fbank features.
For language modeling, the 500 English Byte Pair encoding~(BPE)~\cite{subword_bpe_Kudo_2018} subword units are adopted.
For E2E ASR, we adopt the transformer with the configuration of a 12-layer encoder and a 6-layer decoder ~\cite{attention_is_all_you_need,speech_transformer}, where each self-attention layer has an attention dimension of 512 and 8 heads.
SpecAugment~\cite{park_specaug_2019} is also applied for data augmentation during training.
During decoding, the CTC module is used for score interpolation~\cite{JointCTCAttentionBased_kim_2017} with a weight of 0.3, and a beam width of 10 is applied for beam searching.
All models are built using the ESPnet toolkit~\cite{watanabeESPnetEndtoEndSpeech2018}.

\subsection{Accent Identification and Embedding Extraction}
A pretrained time-delay neural network (TDNN)~\cite{tdnn_2015} based accent identification~(AID) model is used for extracting 256-dimension accent embeddings.
It accepts phone posteriorgram~(PPG) features as input and is trained to predict accent categories.
The accent embeddings are obtained from the penultimate layer output of the AID model.
More details about the AID model can be found in our accent identification system description~\cite{huangAISPEECHSJTUACCENTIDENTIFICATION} for the AESRC 2020 challenge~\cite{AESRC2020Datasetshi}.

\subsection{Exploration of Multi-Basis Adapter Layer}
\label{sec:expr_explore}

We first investigate the performance of the proposed multi-basis adapter layer architecture in Section~\ref{sec:multi_basis} with different injection positions, numbers of bases and types of bases.


\subsubsection{Position of Adapter Layer}
\label{sec:expr_position}

The performance of the baseline model in Section~\ref{sec:baseline} and our proposed models with 4-basis adapter layers are compared in Table~\ref{tab:layer_position}.
Different positions of the adapter layers are evaluated, including $\{1\}, \{6\}, \{12\}, \{1\text{-}6\}$, and $\{1\text{-}12\}$, where $\{m\text{-}n\}$ means injecting adapter layers into the $m^{\text{th}}${\raise.17ex\hbox{$\scriptstyle\sim$}}$n^{\text{th}}$ encoder blocks.

For models with different positions of a single adapter layer injected at only one encoder block (lines~2{\raise.17ex\hbox{$\scriptstyle\sim$}}4), the performance becomes slightly worse as the injection position moves towards the last encoder block.
However, as the number of adapter layers increases, the WER is only on par with single adapter layer-based models.
This indicates that a single adapter layer injected in the first encoder block is already capable of adapting to various accents, while still keeping the parameter efficiency.
Therefore, in the following experiments, only one multi-basis adapter layer is injected in the first encoder block.

\setcounter{table}{0}
\begin{table}[H]
\centering
\caption{
Performance (WER) (\%) comparison of the multi-basis adapter layer positions and numbers.
}
\label{tab:layer_position}
\begin{tabular}{cc | cc cc}
\hline
\multirow{2}[2]{*}{Position} & \multicolumn{2}{c}{Accent} & \multicolumn{2}{c}{Libri} \\
\cmidrule(l{10pt}r{10pt}){2-3} \cmidrule(l{10pt}r{10pt}){4-5} 
& cv & test & dev c/o & test c/o \\
\hline
- & 6.54 & 7.61 & 5.64/11.43 & 6.31/11.68 \\
\{1\} & \textbf{5.91} & \textbf{6.82} & \textbf{{5.17/10.37}} & \textbf{{5.48/10.65}} \\
\{6\} & 5.91 & 6.89 & 5.23/10.41 & 5.51/10.73 \\
\{12\} & 6.08 & 7.08 & 5.20/10.67 & 5.74/10.97 \\
\hline
\{1-6\} & 5.85 & 6.82 & \textbf{5.21/10.39} & 5.65/10.79 \\
\{1-12\} & \textbf{5.82} & \textbf{6.78} & 5.23/10.27 & \textbf{5.61/10.69} \\
\hline
\end{tabular}
\end{table}

\subsubsection{The Number of Bases}
\label{sec:expr_number}

We then explore the impact of different bases numbers (ranging from 2 to 8) on the ASR performance.
As shown in Table~\ref{tab:basis_number}, the WER is gradually decreased as the number of bases increases from 2 to 8.
However, the performance gain is very limited when more than 4 bases are used, but more basis will result in more parameters.
Considering the tradeoff between performance and model size, we adopt the 4-basis adapter layer in our following experiments.

\begin{table}[H]
\centering
\caption{
Performance (WER) (\%) comparison on different numbers of bases in one adapter layer.
}
\label{tab:basis_number}
\begin{tabular}{c cc cc}
\hline
\multirow{2}[2]{*}{\# Bases} & \multicolumn{2}{c}{Accent} & \multicolumn{2}{c}{Libri} \\
 \cmidrule(l{10pt}r{10pt}){2-3} \cmidrule(l{10pt}r{10pt}){4-5} 
 & cv & test & dev c/o & test c/o \\
\hline
- & 6.54 & 7.61 & 5.64/11.43 & 6.31/11.68 \\
2 & 6.23 & 7.34 & 5.36/11.06 & 6.01/11.32 \\
4 & 5.91 & 6.82 & 5.17/\textbf{10.37} & \textbf{5.48/10.65} \\
6 & \textbf{5.89} & \textbf{6.81} & \textbf{5.14}/10.41 & 5.50/10.66 \\
8 & 5.78 & 7.01 & 5.20/10.43 & 5.52/10.71 \\
\hline
\end{tabular}
\end{table}
\setcounter{table}{3}

\begin{table*}[ht]
\centering
\caption{
Performance (WER) (\%) comparison of baseline system and different adaptation methods.
$\adapt_g$ denotes the proposed single-basis accent embedding layer adaptation model in Section~\ref{sec:ebd}, 
and $\adapt_m$ denotes the proposed multi-basis adaptation model introduced in Section~\ref{sec:multi_basis}, with injection only in the first encoder block.
}
\label{tab:final}
\label{tab:full}
\resizebox{\textwidth}{!}{%
\begin{tabular}{ l ccccc ccccc | cc cc}
\hline
\multirow{2}[2]{*}{Model} &  \multicolumn{10}{c}{Accent Test Set} & \multicolumn{2}{c}{Accent} & \multicolumn{2}{c}{Libri} \\
 \cmidrule(l{10pt}r{10pt}){12-13} \cmidrule(l{10pt}r{10pt}){14-15} 
 &  US & UK & IND & CHN & JPN & PT & RU & KR & CAN & ES
& cv & test & dev c/o & test c/o \\
\hline
Baseline 
& 5.75 & 3.17 & 9.32 & 13.49 & 6.66 & 6.38 & 11.04 & 6.80 & 5.21 & 9.88 
& 6.54 & 7.61 & 5.64/11.43 & 6.31/11.68 \\
Finetune 
& 4.92 & 2.82 & 8.34 & 12.00 & 5.92 & 5.66 & 9.78 & 5.82 & 4.25 & 9.18 & 5.85 & 6.83 & 7.84/13.03 & 8.67/13.94 \\
$\adapt_g$
& 5.33 & 3.11 & 8.80 & 12.29 & 6.15 & 6.09 & 9.99 & 6.30 & 4.70 & 9.06
& 5.89 & 6.89  & 5.27/10.39 & 5.76/10.79 \\
$\adapt_m$
& 5.29 & \textbf{2.54} & 8.91 & 11.87 & 6.04 & 5.79 & 9.71 & 6.00 & 4.51 & 8.91
& 5.91 & 6.82 & 5.17/10.37 & 5.48/10.65 \\
$\adapt_g$ + $\adapt_m$
 &\textbf{4.88} & 2.57 & \textbf{8.38} & \textbf{11.54} & \textbf{5.73} & \textbf{5.60} & \textbf{9.71} & \textbf{5.70} & \textbf{4.21} & \textbf{8.51}
& \textbf{5.77} & \textbf{6.68} &  \textbf{4.73/10.22} & \textbf{5.32/10.61} \\
\hline
\end{tabular}%
}%
\end{table*}

\subsubsection{Different Types of Bases}
\label{sec:expr_type}

Table~\ref{tab:module_type} shows the performance of different bases types, including different connection modes (scale, shift, or both scale and shift) in Section~\ref{sec:multi_basis} and different projection module types in the bases.
DNN-based basis uses `Linear` whose encoded dimension is set to 128, while CNN-based basis uses `Conv2d` with a $5\times5$ kernel via 16 channels.
The best performance is achieved when both scaling and shifting are used.
This indicates that the shifting and scaling modes can benefit each other complementarily.
We further test different network types (DNN or CNN) in the bases implementation.
It is observed that CNN-based modules has insufficient ability to extract accent-related information.
In our final system, the DNN-based bases are used for consistency.

\setcounter{table}{2}
\begin{table}[t]
\centering
\caption{
Performance (WER) (\%) comparison of different projection module types and connections in the basis. 
}
\label{tab:module_type}
\resizebox{\columnwidth}{!}{%
\begin{tabular}{cc cc cc}
\hline
\multirow{2}[2]{*}{\tabincell{c}{Network\\Type}} & \multirow{2}[2]{*}{\tabincell{c}{Connection\\Mode}} & \multicolumn{2}{c}{Accent} & \multicolumn{2}{c}{Libri} \\
\cmidrule(l{10pt}r{10pt}){3-4} \cmidrule(l{10pt}r{10pt}){5-6} 
 & & cv & test & dev c/o & test c/o \\
\hline
- & - & 6.54 & 7.61 & 5.64/11.43 & 6.31/11.68 \\
DNN & shifting-only & 5.96 & 6.91 & 5.22/10.41 & 5.51/10.78 \\
DNN & scaling-only & 5.95 & 6.99 & 5.23/10.44 & \textbf{5.46}/10.70 \\
DNN & both & \textbf{5.91} & \textbf{6.82} & \textbf{5.17/10.37} & 5.48/\textbf{10.65} \\
CNN & shifting-only & 6.12 & 7.11 & 5.16/10.61 & 5.67/11.10 \\
\hline
\end{tabular}%
}%
\end{table}


%

\subsection{Results Comparison of Different Adaptation Methods}
\label{sec:expr_full}

In this section, we present the detailed performance comparison of all proposed models and the baselines in Table~\ref{tab:full}.
Fine-tuning the baseline model for accented data is an intuitive way to perform adaptation on accented data, which is shown in the second line of Table~\ref{tab:full}. 
However, this is not feasible for some unseen accents like Spain~(ES), which is unavoidable during inference. On the other aspect, it degrades the performance on standard data, i.e.~Librispeech evaluation sets.
The gated adapter layer in Section~\ref{sec:ebd} is denoted as $\adapt_g$ in the table, which shows significant improvement on both Librispeech and accent datasets.
Denote by $\adapt_m$ the proposed multi-basis adapter layer introduced in Section~\ref{sec:multi_basis}, adapter layer $\adapt_m$ is injected only in the first encoder block, which consists of 4 bases that are structured by DNN-based projection modules.
Furthermore, we combine $\adapt_g$ and $\adapt_m$ by computing the output as $h^i + \adapt_m(h^i + \adapt_g(h^i, z), z)$.
We observe that the final proposed method $\adapt_g$ + $\adapt_m$ consistently outperforms the baseline model which shows that the proposed method can learn accent-related information effectively and improve the robustness of speech recognition against accent variation.

\subsection{Visualization of Multi-Basis Adapter Layer}
\label{sec:expr_visual}

\begin{figure}[h]
\centering
\begin{subfigure}{0.2\textwidth}
\includegraphics[width=1.0\linewidth]{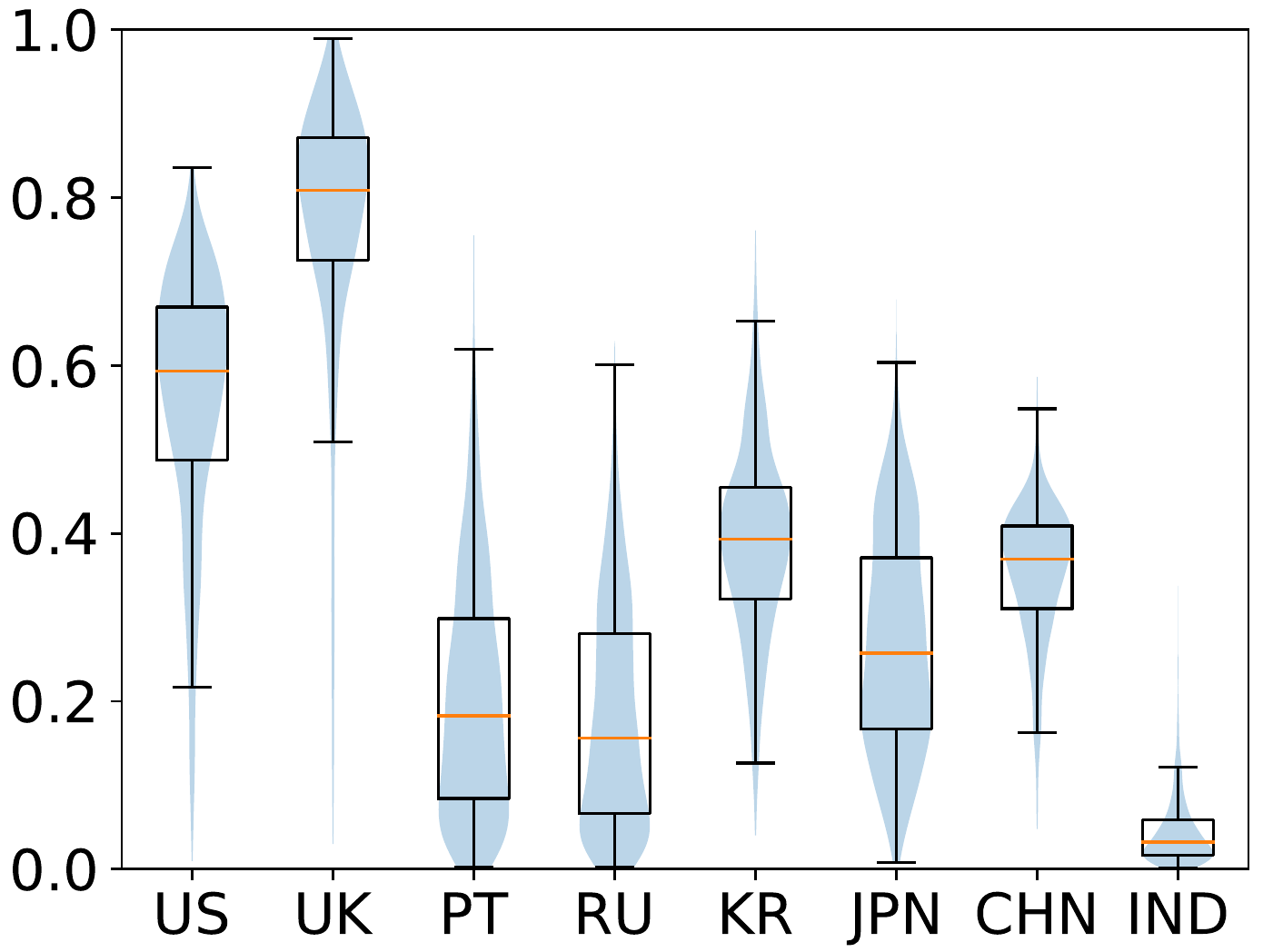}
\caption{Basis One}
\end{subfigure}
\begin{subfigure}{0.2\textwidth}
\includegraphics[width=1.0\linewidth]{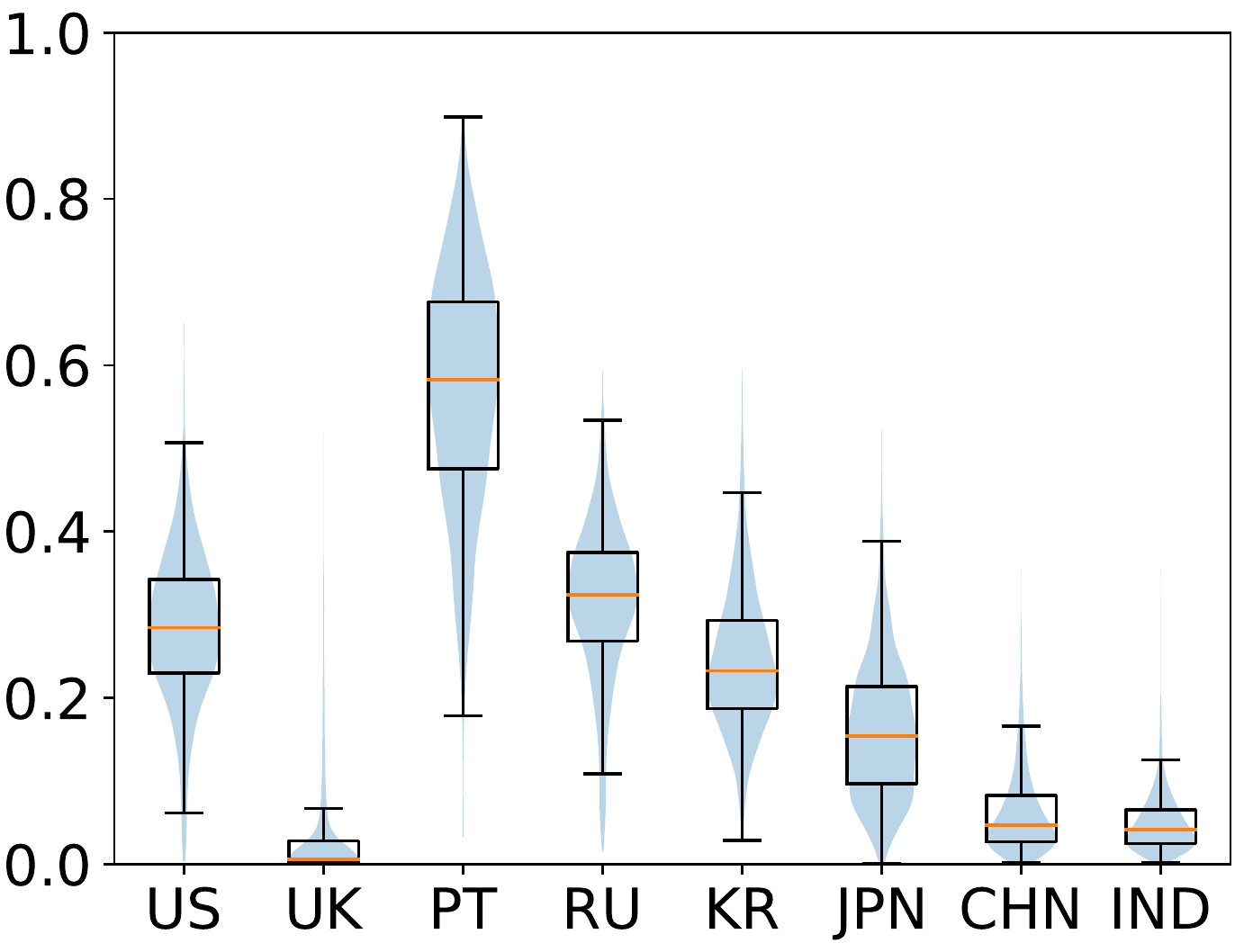}
\caption{Basis Two}
\end{subfigure}
\begin{subfigure}{0.2\textwidth}
\includegraphics[width=1.0\linewidth]{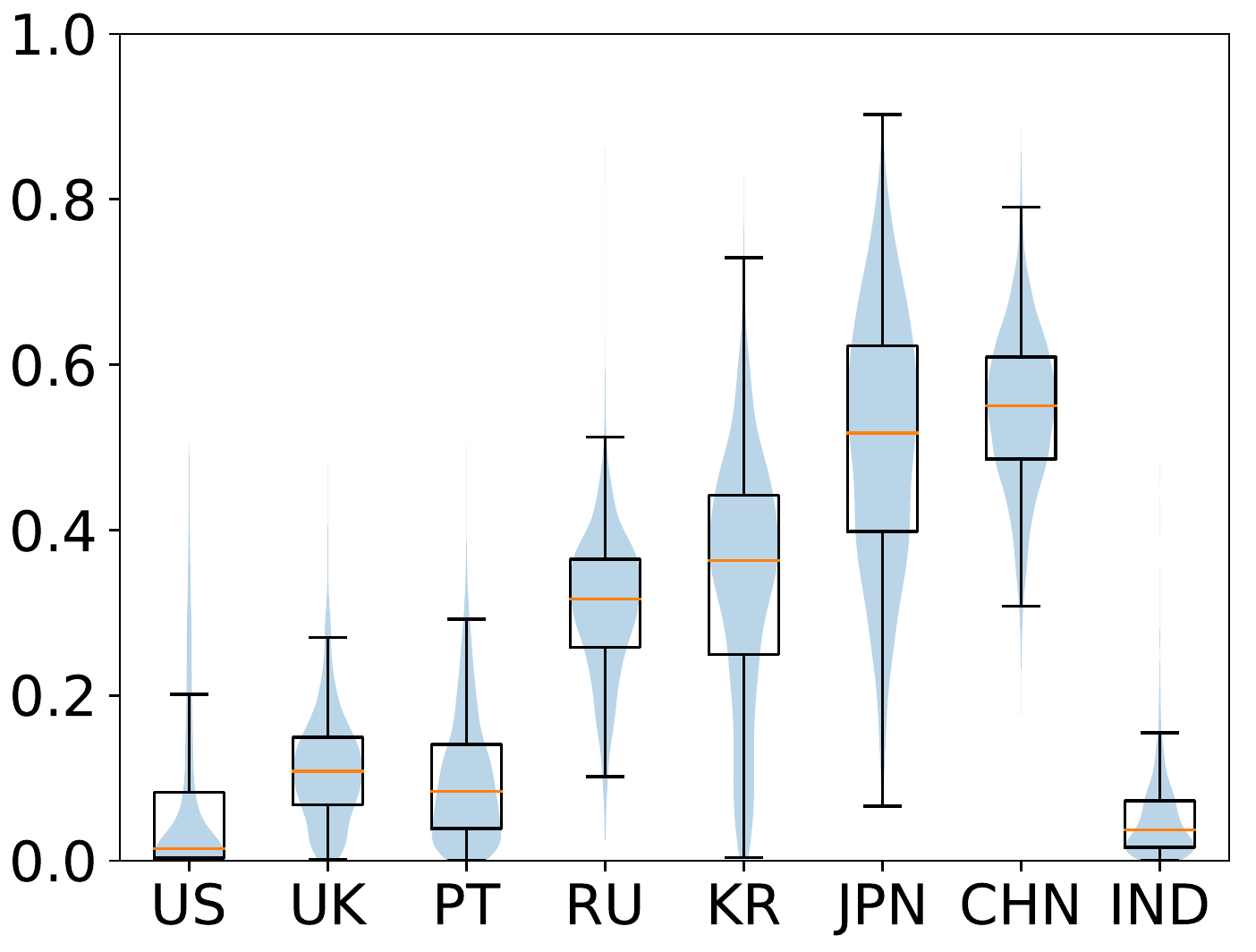}
\caption{Basis Three}
\end{subfigure}
\begin{subfigure}{0.2\textwidth}
\includegraphics[width=1.0\linewidth]{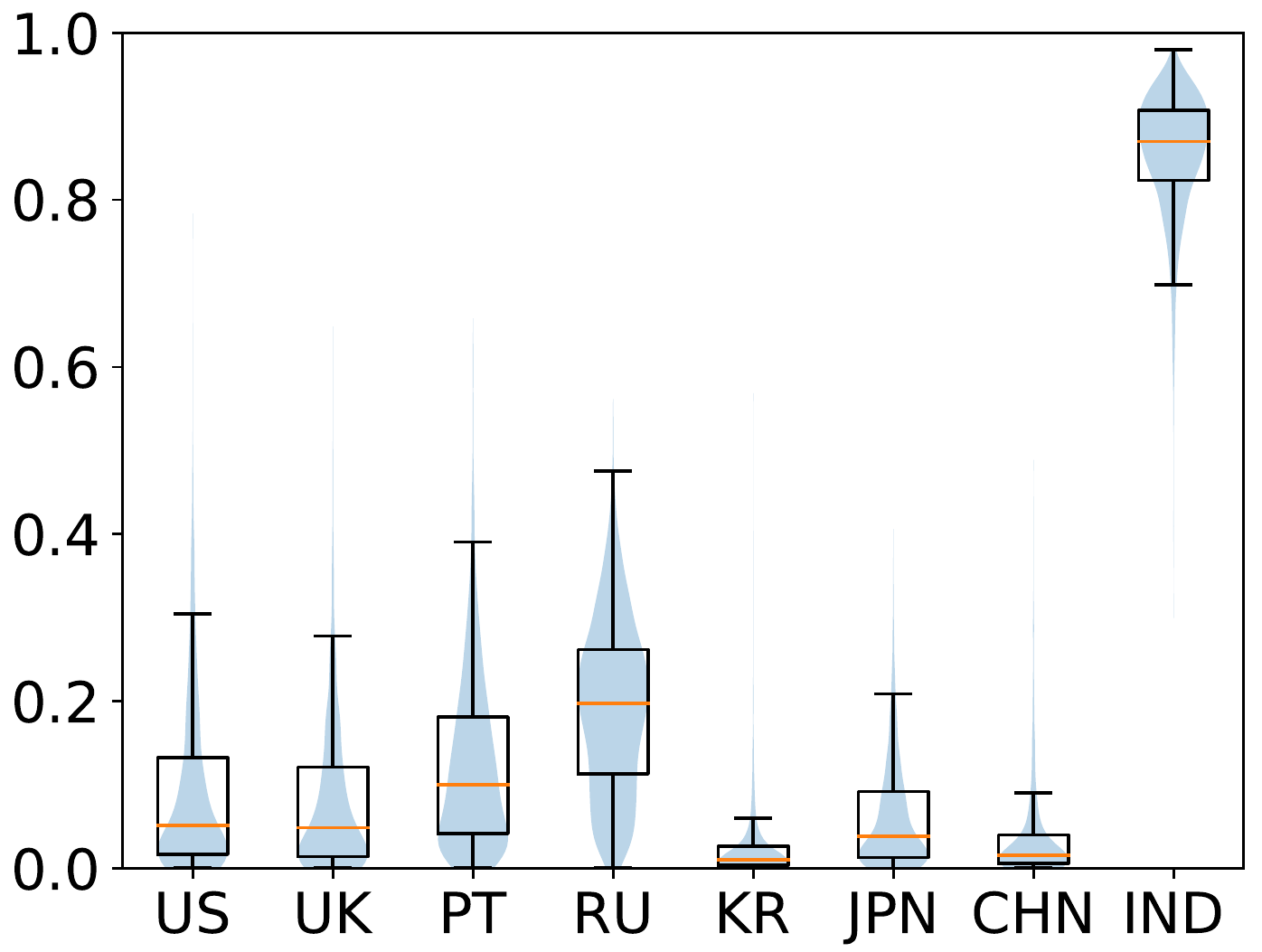}
\caption{Basis Four}
\end{subfigure}
\caption{
Boxplot and violinplot visualization of interpolation coefficent distributions for each basis.
The vertical axis shows the interpolation coefficent $\alpha_i$, where $i$ is the basis index.
The horizontal axis is the accent categories.
}
\label{fig:interpolation}
\end{figure}

Figure~\ref{fig:interpolation} shows the coefficient distributions on each basis from the 4-basis adapter layer model.
Accents with large coefficients in each basis are assumed to be more correlated to that basis.
It can be clearly seen that different bases capture a different set of highly correlated accents.
For example, \textit{Basis Two} focuses mostly on extracting information about the Portuguese~(PT) accent, and then the American~(US) and Russian~(RU) accents.
The inherent correlation between different accents can be also revealed from this figure.
For example, American~(US) and British~(UK) accents have consistently high correlations with \textit{Basis One}, and much lower correlations with other bases.
Meanwhile, Indian~(IND) and Japanese~(JPN) accents have distinct preferences for bases: IND accent prefers \textit{Basis Four} while JPN accent prefers \textit{Basis Three}.
Results demonstrate that our proposed multi-basis adapter layer approaches can well-capture the accent-dependent information with the guidance of accent embeddings, thus improving the multi-accent ASR performance.

\section{Conclusions}
\label{sec:conclusions}
In this paper, we explore a layer-wise adapter architecture to improve E2E-based multi-accent speech recognition models.
The proposed models transform accent-unrelated input into an accent-related space by injecting small adapter layers in the ASR encoder blocks.
The models use a pretrained accent identification network for accent embeddings estimation, a shared predictor for learning interpolation coefficients of different adapter bases, and several accent-related bases for accent adaptation.
Experimental results reveal that we outperform the baseline model up to 12\% relative WER reduction on AESRC2020 cv/test sets and 10\% relative WER reduction on Librispeech dev/test sets as well.
In future work, we would like to investigate different combination methods between the accent embedding and acoustic features, i.e.~the internal structures of the adapter basis.

\section{Acknowledgements}

This work was supported by the China NSFC projects~(No. 62071288 and No. U1736202).
Experiments have been carried out on the PI supercomputers at Shang-hai Jiao Tong University.

\bibliographystyle{IEEEtran}

\bibliography{output}

\end{document}